\documentclass[hyper]{JHEP} 

\usepackage{epsfig}

\newcommand\fverb{\setbox\pippobox=\hbox\bgroup\verb}

\newcommand\fverbdo{\egroup\medskip\noindent%

            \fbox{\unhbox\pippobox}\ }

\newcommand\fverbit{\egroup\item[\fbox{\unhbox\pippobox}]}

\newbox\pippobox

\title{T-Duality For String in Ho\v{r}ava-Lifshitz Gravity}
\author{J. Kluso\v{n}\\
Department of
Theoretical Physics and Astrophysics\\
Faculty of Science, Masaryk University\\
Kotl\'{a}\v{r}sk\'{a} 2, 611 37, Brno\\
Czech Republic\\
\vskip 5mm
{\bf and} \\
\vskip 5mm
Theory Division, Physics Department, CERN, \\
CH-1211 Geneva 23, Switzerland\\
E-mail: \email{klu@physics.muni.cz} \ , }
\author{Kamal L. Panigrahi\\
Department of Physics and Meteorology \&\\
Centre for Theoretical Studies,\\
Indian Institute of Technology Kharagpur,\\
Kharagpur- 721 302, INDIA\\
E-mail: \email{panigrahi@phy.iitkgp.ernet.in}}
\preprint{CERN-PH-TH/2010-144
}
\abstract{We continue our study of the
Lorentz breaking string theories. These theories are defined
as string theory with modified Hamiltonian
constraint which breaks the
Lorentz symmetry of target space-time.
 We analyze
properties of this theory in the target space-time that
possesses isometry along one direction. We
also derive the T-duality rules
for Lorentz breaking string theories
 and show that they are the same as that of Buscher's T-duality for the
relativistic strings.}

\def\tV{\tilde{V}}
\def\bV{\mathbf{V}}
\def\hV{\hat{V}}

\def\mH{\mathcal{H}}

\def\hN{\hat{N}}

\def\partt{\partial_\tau}
\def\parts{\partial_\sigma}

\def\bx{\mathbf{x}}

\newcommand{\hg}{\hat{g}}

\newcommand{\hh}{\hat{h}}

\newcommand{\tphi}{\tilde{\phi}}
\newcommand{\mL}{\mathcal{L}}

\def\pb #1{\left\{#1\right\}}

\begin{document}
\section{Introduction and Summary}\label{first}
The recent proposal of Ho\v{r}ava \cite{Horava:2009uw,Horava:2008ih,
Horava:2008jf}
for a candidate theory of gravity
that is symmetric under the Lifshitz type of anisotropy scaling of space-time
coordinates $t \rightarrow l^{z} t , \>\>\>\>
x^i \rightarrow l x^i , $ where $z$ being the scaling exponent, has been
a very interesting area of research since the last year
\footnote{Some aspects of Ho\v{r}ava-Lifshitz theory has been discussed
in for example \cite{Tang:2009bu}- \cite{Ghodsi:2009rv}}.
This theory is constructed as a UV completion of Hilbert-Einstein
gravity so that it is perturbatively renormalizable.
This modification is possible only when
we sacrifice Lorentz symmetry at high energy. However it is again
observable at low energy. Among various versions of  Ho\v{r}ava-Lifshitz
theories, only the class of projectable theories where the
so called lapse function
depends only on time, is the consistent choice. It is often, as in the present
case, interesting to study theories with broken general covariance. The
so called Lorentz symmetry breaking Hamiltonian formalism has been used to
study the point particles and strings in  Ho\v{r}ava gravity. The basic
idea of this Lorentz breaking Hamiltonian formalism is that time and
spatial components of momenta are treated differently. Indeed in
\cite{Kluson:2010aw} the construction of  a new string theory, called
Lorentz-breaking string theory (LBS) has been studied extensively by
generalizing the point particle dynamics
\cite{Capasso:2009fh,Suyama:2009vy,Romero:2009qs,Mosaffa:2010ym,
Kiritsis:2009rx,Rama:2009xc}
in Ho\v{r}ava-Lifshitz gravity.
The basic idea of the construction of the LBS theory is following. We
start with the Hamiltonian formulation of two dimensional theory when
the Hamiltonian is linear combination of two constraints: the spatial
diffeomorphism constraint and Hamiltonian constraint. As opposite
to the Hamiltonian formulation of Polyakov string we consider the
Hamiltonian constraint that breaks the Lorentz invariance of the
target space-time in the similar way as in the point particle case
\cite{Capasso:2009fh,Suyama:2009vy,Romero:2009qs,Mosaffa:2010ym,
Kiritsis:2009rx,Rama:2009xc}. However as opposite to the point particle
case now the world-sheet modes depend on spatial coordinate of
world-sheet theory so that it is possible to define many
LBS theories that reduce to the point particle Hamiltonian
constraint \cite{Capasso:2009fh,Suyama:2009vy,Romero:2009qs,Mosaffa:2010ym,
Kiritsis:2009rx,Rama:2009xc} in case of the world-sheet
 dimensional reduction. In doing so, the consistency of the LBS theory
demands that the spatial component of the world-sheet metric
has to be dynamical.
As a by product, the world-sheet theory is no more invariant under full
two dimensional diffeomorphism but only under the world-sheet foliation
preserving transformation. Furthermore, the consistency of the Hamiltonian
dynamics of LBS theory implies that the world-sheet lapse has to obey
the projectability condition consistent with the
of Ho\v{r}ava-Lifshitz gravity.
 On the other hand it is natural
 to demand that the Hamiltonian constraint of LBS theory
 reduces to the Hamiltonian constraint of the relativistic
 string in the case when the target Ho\v{r}ava-Lifshitz gravity
reduces to  General Relativity in low energy regime.
This requirement now implies
 that we should consider LBS theory where the world-sheet mode
  $x^0$ depends on
the  world-sheet spatial coordinate $\sigma$ as well which is  more
 general situation than was consider in the paper   \cite{Kluson:2010aw}.
Then we will be able to show  that the Hamiltonian
constraint  reduces to the
Hamiltonian constraint for
the relativistic string in the  limit which has
been used to recover General Theory of Relativity from
Ho\v{r}ava gravity. However at this place we should stress
one important point that makes the construction
of LBS theory as intricate as the construction of
Ho\v{r}ava-Lifshitz gravity. Explicitly, we argue in
\cite{Kluson:2010aw} that the consistency of the
Hamiltonian formulation of LBS theory forces us to
consider the world-sheet lapse function that depends
on the world-sheet time coordinate $\tau$ only. As a result the LBS theory reduces to the
Polyakov action when however the lapse function
does not depend on the world-sheet spatial coordinate.
In other words LBS theory does not reduce to the Polyakov
string in the IR limit of target Ho\v{r}ava-Lifshitz gravity.

Despite of this fact, we feel that it is interesting to
study LBS theory further as a toy model of the theory
with broken Lorentz invariance that is more general then
the corresponding point particle action. We discussed
 the symmetries of
the action and have shown that the action is invariant under the target space
foliation preserving diffeomorphism
and under world-sheet foliation preserving diffeomorphism
\cite{Kluson:2010aw}. We further derive
the T-duality rules for LBS  string
 and show that they are same as that of Buscher's T-duality for the
relativistic strings\cite{Buscher:1985kb,Buscher:1987sk,Buscher:1987qj}.
It would be interesting to study other extended
objects like D-branes in LBS theory. In particular it will be desired to
look for D-brane action in Ho\v{r}ava-Lifshitz background
and examine their fate by using the T-duality
transformation.

The rest of the paper is organized as follows. In section-2, we generalize the
construction of LBS theory of \cite{Kluson:2010aw} to include
other auxiliary fields and more general world sheet modes.
In section-3, we present the symmetries of the LBS theory action and show
they are invariant under the target space foliation preserving
diffeomorphism transformation. Finally, section-4 is devoted to
the study of T-duality transformation of the LBS theory.

\section{Review of LBS Theory}
 In this section we review and slightly generalize
the construction of LBS theory
given in \cite{Kluson:2010aw} where more
details of motivation for this construction can be found.
As in \cite{Kluson:2010aw}
we begin with the following Hamiltonian formulation
of LBS \begin{eqnarray}\label{defH}
H&=&\int_{\Sigma} d\sigma \mH(\sigma) \ , \quad
\mH(\sigma)=n_\tau(\tau) \mH_\tau(\sigma)
+ n^\sigma(\sigma)
\mH_\sigma(\sigma)+\nonumber \\
&+& \lambda_\tau(\tau)
\pi^\tau(\tau)+\lambda_\sigma(\sigma) \pi^\sigma(\sigma)+
v_A(\sigma) P_A(\sigma)+v_B(\sigma) P_B(\sigma) \ , \nonumber \\
\end{eqnarray}
where
\begin{eqnarray}\label{defHtausigma}
\mH_\tau&=&-\pi\alpha' \frac{1}{\sqrt{\omega}
N^2}
(p_0-N^ip_i)^2+\sqrt{\omega}
G\left(-\frac{1}{4\pi\alpha'\omega}
N^2\parts x^0 \parts x^0\right) +\nonumber
\\
&+&B\left( \pi\alpha' \frac{1}{\omega}
p_i h^{ij}p_j+
\frac{(z-1)}{2\sqrt{\omega}}
\pi^\omega \omega^2 \pi^\omega +\right. \nonumber \\
&+& \left.\frac{1}
{4\pi\alpha'}\frac{1}{\omega}
(\parts x^i+N^i\parts x^0)
h_{ij}(\parts x^j+N^j\parts x^0)-A\right)+\sqrt{\omega}F(A) \
\ , \nonumber \\
\mH_\sigma&=& p_M\parts x^M- 2\omega
\nabla_\sigma \pi^\sigma
\ , \nonumber \\
\end{eqnarray}
where $x^M, M,N=0,\dots,D$ are world-sheet
modes that parameterize the embedding of the string
 into
target space-time $\mathcal{M}$ with general
metric $g_{MN}$
and where $p_M$ are conjugate
momenta with following non-zero Poisson brackets
\begin{equation}
\pb{x^M(\sigma),p_N(\sigma')}=
\delta^M_N\delta(\sigma-\sigma') \ .
\end{equation}
Further,  we
introduced two dimensional metric $\gamma_{\mu\nu}$
in $1+1$ formalism
\begin{equation}\label{gamma}
\gamma_{\alpha\beta}= \left(
\begin{array}{cc}
-n^2_\tau+ \frac{1}{\omega
}n_\sigma^2 &
n_\sigma \\
 n_\sigma & \omega \\
\end{array}\right) \ ,
\end{equation}
where $n_\tau$ is a world-sheet lapse,
$n_\sigma$ is  a world-sheet shift,
$ n^\sigma\equiv \frac{n_\sigma}{\omega}$
 and
$\omega$ is a spatial part of world-sheet metric
and where $\pi^\tau,\pi^\sigma$ and $\pi^\omega$
are corresponding conjugate momenta with following
non-zero Poisson brackets
\begin{equation}
\pb{n_\tau,\pi^\tau}=1 \ ,
\quad \pb{n_\sigma(\sigma),\pi^\sigma(\sigma')}=
\delta(\sigma-\sigma') \ , \quad
\pb{\omega(\sigma),\pi^\omega(\sigma')}=
\delta(\sigma-\sigma') \ .
\end{equation}
We further defined world-sheet covariant
derivative
\begin{equation}
\nabla_\sigma n_\sigma=\parts
n_\sigma-\Gamma n_\sigma\ , \quad
\nabla_\sigma \pi^\omega=
\parts \pi^\omega+\Gamma \pi^\omega \ , \quad
\Gamma=\frac{1}{2\omega}\parts \omega
 \ .
\end{equation}
We  consider  target space-time
$\mathcal{M}$ labeled with coordinates
$t=x^0,\bx=(x^1,\dots,x^D)$ with the metric
in $1+1$  form
\begin{equation}
g_{00}=-N^2+N_ih^{ij}N_j \ , \quad
g_{0i}=N_i \ , \quad g_{ij}=h_{ij} \ ,
\quad  \det g=-N^2\det h \
\end{equation}
with inverse
\begin{equation}
g^{00}=-\frac{1}{N^2} \ , \quad
g^{0i}=\frac{N^i}{N} \ , \quad
g^{ij}=h^{ij}-\frac{N^i N^j}{N^2} \ .
\end{equation}
Note that the dynamics of target space metric $g_{MN}$
is governed by Ho\v{r}ava-Lifshitz gravity action.
 Further, $A,B$ and corresponding
conjugate momenta $P_A,P_B$ are auxiliary modes
 with Poisson brackets
\begin{equation}
\pb{A(\sigma),P_A(\sigma')}=
\delta(\sigma-\sigma') \ , \quad
\pb{B(\sigma),P_B(\sigma')}=
\delta(\sigma-\sigma') \ .
\end{equation}
Finally  $\lambda_\tau,\lambda_\sigma, v_A,v_B$ are Lagrange multipliers
that ensure that $\pi^\tau\approx 0 \ , \pi^\sigma(\sigma)\approx 0 \ ,
P_A(\sigma)\approx 0 \  ,  P_B(\sigma)\approx 0$ are
primary constraints of the theory.  Note that following arguments
given in \cite{Kluson:2010aw} we presume that
$n_\tau$ depends on $\tau$ only.
Then
the requirement of the preservation of the primary constraints
implies following secondary ones
\begin{eqnarray}
\partt \pi_\tau&=&\pb{\pi_\tau,H}=
-\int d\sigma \mH_\tau\approx 0 \ , \nonumber \\
\partt \pi^\sigma&=&\pb{\pi^\sigma,H}=
-\frac{1}{\omega}\mH_\sigma\approx 0 \ , \nonumber \\
\partt P_A&=&\pb{P_A,H}=B-\sqrt{\omega}F'(A)\equiv
G_A\approx 0 \ , \nonumber \\
\partt P_B&=&\pb{P_B,H}=-
\left( \pi\alpha' \frac{1}{\omega}
p_i h^{ij}p_j+
\frac{(z-1)}{2\sqrt{\omega}}
\pi^\omega \omega^2 \pi^\omega +\right. \nonumber \\
&+& \left.\frac{1}
{4\pi\alpha'}\frac{1}{\omega}
(\parts x^i+N^i\parts x^0)
h_{ij}(\parts x^j+N^j\parts x^0)-A\right)\equiv G_B\approx 0 \ .
\nonumber \\
\end{eqnarray}
It is easy to show that the secondary constraints $G_A\approx 0,G_B\approx 0$
 together with the primary ones  $P_A\approx 0, P_B\approx 0$ form
the collection of the second class constraints. Solving the
system of the second class constraints
$(P_A,P_B,G_A,G_B)$ we express $A,B$ as
functions of the canonical variables $x^M,p_M$
and we find non-linear form of the Hamiltonian
constraint. This procedure was extensively studied
in  \cite{Kluson:2010aw} so that we skip
all details and recommend an interesting reader to
look at this reference.

As opposite to the case of the Hamiltonian
formulation of the relativistic string
we see that the Hamiltonian constraint
(\ref{defHtausigma}) contains
 kinetic term for  $\pi^\omega$.
 Arguments why
there is non-trivial dynamics the
spatial part of the metric $\omega$
were given in \cite{Kluson:2010aw}
and we briefly recapitulate here.
Let us imagine for the time
being  that $\pi^\omega$  is a primary constraint
of the theory. However we see from
(\ref{defH}) that the Hamiltonian constraint
depends on
$\omega$ in non-trivial way. Then if
$\pi^\omega\approx 0$ were be the primary
constraint of the theory  we would find
overconstrained theory due to the
requirement of the consistency of this
constraint with the time evolution of
the system. Then in order to avoid
imposing additional constraint on the
system we demand
 that $\omega$ is a dynamical
mode with kinetic term in the action.

We see that  the Hamiltonian  constraint
(\ref{defHtausigma}) is characterized  by
the presence of two functions $F$
and $G$.  As in \cite{Capasso:2009fh,Suyama:2009vy,Romero:2009qs,Mosaffa:2010ym,
Kiritsis:2009rx,Rama:2009xc}
we presume that $F(A)$ has
the form $F(A)=A+
\sum_{n=2}^{z} \lambda_n A^n$ with
$z$ being the critical exponent of the
Ho\v{r}ava-Lifshitz gravity. It is
believed that in the IR limit  the
Ho\v{r}ava-Lifshitz gravity reduces
to the ordinary General Relativity
when $z=1$. Let us now study properties
of the Hamiltonian constraint
(\ref{defHtausigma}) in this limit. Firstly
we see  that the kinetic
term for the spatial part of the
metric vanishes for $z\rightarrow 1$.
Since  we consider more general
case than in \cite{Kluson:2010aw}
when $x^0$ depends on $\sigma$ we
mean that  it is natural to add
into (\ref{defHtausigma}) the term
$G(-\frac{1}{4\pi \alpha'\omega^2}N^2
\parts x^0\parts x^0)$.
We  assume that
 $G(A)=A+\sum_{n=2}^z \omega_n A^n$
where $\omega_n$ are constants.
Then it is easy to see that in the limit $
z\rightarrow 1$ we have $F(A)\rightarrow 1 \ ,
\quad G(A)\rightarrow 1$ and hence
we find that for $z\rightarrow 1$ the Hamiltonian
constraint given in  (\ref{defHtausigma})
takes the following form
\begin{eqnarray}\label{Hamconflat}
\mH_\tau&=&-\pi\alpha' \frac{1}{\sqrt{\omega}
N^2}
(p_0-N^ip_i)^2
-\frac{1}{4\pi\alpha'\sqrt{\omega}}
N^2\parts x^0 \parts x^0 +\nonumber
\\
&+&B\left( \pi\alpha' \frac{1}{\omega}
p_i h^{ij}p_j+
\frac{1}
{4\pi\alpha'}\frac{1}{\omega}
(\parts x^i+N^i\parts x^0)
h_{ij}(\parts x^j+N^j\parts x^0)-A\right)+\sqrt{\omega}A \
\ . \nonumber \\
\end{eqnarray}
Solving now the second class constraints
$P_A,P_B,G_A,G_B$ is equivalent to the
integration out  $A$ and $B$ from
(\ref{Hamconflat}) and we find
that the Hamiltonian constraint
(\ref{Hamconflat}) reduces to
\begin{eqnarray}
\mH_\tau&=&-\pi\alpha' \frac{1}{\sqrt{\omega}
N^2}
(p_0-N^ip_i)^2+
 \pi\alpha' \frac{1}{\sqrt{\omega}}
p_i h^{ij}p_j-\nonumber \\
&-&\frac{1}{4\pi\alpha'\sqrt{\omega}}
N^2\parts x^0 \parts x^0 +
\frac{1}
{4\pi\alpha'}\frac{1}{\sqrt{\omega}}
(\parts x^i+N^i\parts x^0)
h_{ij}(\parts x^j+N^j\parts x^0)=
 \nonumber \\
&=& -\pi\alpha' \frac{1}{\sqrt{\omega}}
p_N g^{MN}p_N
-\frac{1}{4\pi\alpha'\sqrt{\omega}}
\partial_\sigma x^M g_{MN}\partial_\sigma x^N \ .
\nonumber \\
\end{eqnarray}
This is clearly the Hamiltonian constraint
of the Polyakov action. We see
 that the LBS action reduces to the Polyakov
in the IR limit of target space-time.
However we should
stress the crucial point in the formulation
of LBS theory that is the same way resembles
the same problem as in  the
formulation of the Ho\v{r}ava-Lifshitz gravity.
Explicitly, we argued in
\cite{Kluson:2010aw}
that LBS  theory is
well defined only when the lapse function
$n$ depends on $\tau$ only i.e. $n=n(\tau)$. However
then we see that in the IR limit of the
Ho\v{r}ava-Lifshitz gravity the LBS theory
reduces to the Polyakov action where
$\gamma_{00}=-n_\tau$ depends on $\tau$ only
and hence the full diffeomorphism invariance is
not restored.


The next step is to find Lagrangian
corresponding to Hamiltonian
(\ref{defH}).  To do this we determine
the time derivatives of $x^M,A,B,\omega$
\begin{eqnarray}
\partt x^0&=&\pb{x^0,H}=
-\frac{2\pi\alpha'}{N^2\sqrt{\omega}}(p_0-N^i p_i)
n_\tau +n^\sigma \parts x^0 \ , \nonumber \\
\partt
x^i&=& \frac{2\pi\alpha'}{N^2\sqrt{\omega}
}N^i(p_0-N^ip_i)n_\tau+
2\pi\alpha' B
\frac{1}{\omega}
h^{ij}p_jn_\tau+n^\sigma
\parts x^i \ ,  \nonumber \\
\partt A&=&\pb{A,H}=v_A \ , \quad
\partt B=\pb{B,H}=v_B \ , \nonumber \\
\partt \omega&=&\pb{\omega,H}=
n_\tau \frac{B(z-1)\omega^2}{\sqrt{\omega}}\pi^\omega
+2\nabla_\sigma n_\sigma \ .\nonumber \\
\end{eqnarray}
It is convenient to introduce following
object
\begin{eqnarray}
K_\sigma&=&\frac{1}{n_\tau} (\partt
\omega-2\nabla_\sigma n_\sigma) \ .
\nonumber \\
 \end{eqnarray}
Then it is simple task to find corresponding
Lagrangian
\begin{eqnarray}\label{mlAB}
\mL&=& p_M \partt x^M+\partt Ap_A+\partt B p_B+
\partt \omega \pi^\omega-\mH= \nonumber \\
&=&-\frac{\sqrt{\omega}}{4\pi\alpha'}\frac{1}{n_\tau}
(\partt x^0-n^\sigma\parts
x^0)^2-n_\tau \sqrt{\omega} G\left(
-\frac{1}{2\pi\alpha'\omega} N^2\parts x^0
\parts x^0\right)+
\nonumber \\
&+&\omega n_\tau \frac{1}{B}
\left(\frac{1}{4\pi\alpha'}\frac{1}{n^2_\tau}
(V^i_\tau-n^\sigma V^i_\sigma)
 h_{ij}(V^j_\tau-n^\sigma V^j_\sigma)
 +\frac{1}{2(z-1)}K_\sigma \frac{1}{\omega^2}K_\sigma\right)-
 \nonumber \\
 &-& Bn_\tau\left(\frac{1}{2\pi\alpha'\omega}
 V^i_\sigma h_{ij}V^j_\sigma-A\right)-\sqrt{\omega} n_\tau F(A) \ ,
 \nonumber \\
\end{eqnarray}
where
\begin{equation}
V_\tau^i=\partt x^i+N^i\partt x^0 \ ,
\quad
V_\sigma^i=\parts x^i+N^i\parts x^0 \ .
\end{equation}
Observe that this theory is well defined in
case when $z\rightarrow 1$ on condition that
$K_\sigma=0$ which is in agreement with our
requirement that this theory reduces to
Polyakov action in this limit with  exception
that $\gamma_{00}$ depends on $\tau$ only.

Finally we  integrate out  $A$ and
$B$ from (\ref{mlAB}). The equation of motion for $A$
implies
\begin{equation}\label{BA}
B-\sqrt{\omega}F'(A)=0
\end{equation}
while the equation of motion for $B$
implies
\begin{eqnarray}\label{eqA}
&-&\frac{\omega }{B^2}
\left( \frac{1}{4\pi\alpha' n_\tau^2}
(V_\tau^i-n^\sigma V^i_\sigma) h_{ij}
(V_\tau^j-n^\sigma V^j_\sigma)
+\frac{1}{2(z-1)}K_\sigma \frac{1}{\omega^2}
K_\sigma\right)
 -\nonumber \\
 &-&
\left(\frac{1}{2\pi\alpha'\omega}
V^i_\sigma h_{ij}V^j_\sigma-A\right)=0 \ .
\nonumber \\
\end{eqnarray}
Inserting (\ref{BA}) into (\ref{eqA})
we find the equation for $A$ in the form
\begin{eqnarray}
&-&\frac{1}{F'^2(A)
}\left(\frac{1}{4\pi\alpha' n_\tau^2}
(V_\tau^i-n^\sigma V^i_\sigma) h_{ij}
(V_\tau^j-n^\sigma V^j_\sigma)
+\frac{1}{2(z-1)}K_\sigma \frac{1}{\omega^2}
K_\sigma\right)-\nonumber \\
 &-&
\left(\frac{1}{2\pi\alpha'}
V^i_\sigma h_{ij}V^j_\sigma-A\right)=0 \
\nonumber \\
\end{eqnarray}
that in principle allows as to find
$A$ as
\begin{equation}
A=\Psi\left( \frac{1}{4\pi\alpha'n_\tau^2}
(V^i_\tau-n^\sigma V^i_\sigma)
h_{ij}(V^j_\tau-n^\sigma V^j_\sigma)+
\frac{1}{2(z-1)}K_\sigma \frac{1}{\omega^2}
K_\sigma,
\frac{1}{2\pi\alpha'\omega} V_\sigma^i
h_{ij}V^j_\sigma\right) \ .
\end{equation}
Collecting all these results together
we find the Lagrangian density in the form
\begin{eqnarray}\label{mLLBS}
\mL&=&\sqrt{\omega}n_\tau\left[
-\frac{1}{4\pi\alpha'}\frac{1}{n_\tau^2}
(\partt x^0-n^\sigma\parts
x^0)^2- G\left(-\frac{1}{2\pi\alpha'\omega} N^2\parts x^0
\parts x^0\right)-\right.
\nonumber \\
&-& \left.  F'(\Psi)\left(\frac{1}{2\pi\alpha'\omega}
 V^i_\sigma h_{ij}V^j_\sigma-A\right)-2 F(\Psi) \right] \  .
 \nonumber \\
\end{eqnarray}
This it the final form of the Lorentz breaking
string theory Lagrangian. In the next section we
study invariance of the action $S=\int d\tau d\sigma
\mL$ under local and global world-sheet symmetries.
\section{Symmetries of the LBS Action}
 We start with the global
transformations  from the point
of view of the string world-sheet
theory. These transformations
correspond to the foliation preserving
diffeomorphism of the target space-time
\cite{Horava:2009uw,Horava:2008ih}
\begin{eqnarray}\label{deffpdts}
x'^0(\tau,\sigma)&=&x^0(\tau,\sigma)+f(x^0(\tau,\sigma))
\ ,
\nonumber \\
x'^i(\tau,\sigma)&=&
x^i(\tau,\sigma)+\zeta^i(\tau,\sigma) \ ,
\nonumber \\
\end{eqnarray}
where $f(x^0),\zeta^i(x^0,\bx)$ are infinitesimal parameters.
Note that under these transformations
the metric component transform as
\begin{eqnarray}\label{Ntr}
N'_i(x'^0,\bx')&=& N_i(x^0,\bx)
-N_i(x^0,\bx)
\dot{f}(x^0)-N_j(x^0,\bx)\partial_i
\zeta^j(x^0,\bx)-g_{ij}(x^0,\bx)
\dot{\zeta}^j(t,\bx) \ ,   \nonumber \\
 N'^i(x'^0,\bx')
&=&N^i(x^0,\bx)+N^j(x^0,\bx)\partial_j
\zeta^i(x^0,\bx)-
N^i(x^0,\bx)\dot{f}-\dot{\zeta}^i(x^0,\bx)
\ ,
\nonumber \\
N'(x'^0)&=&N(x^0)-N(x^0) \dot{f}(x^0) \  \nonumber \\
\end{eqnarray}
and
\begin{eqnarray}\label{trm}
g'_{ij}(x'^0,\bx')&=&g_{ij}(x^0,\bx)-
g_{il}(x^0,\bx)\partial_j
\zeta^l(x^0,\bx)-\partial_i
\zeta^k(x^0,\bx) g_{kj}(x^0,\bx) \ , \nonumber \\
g'^{ij}(x'^0,\bx')&=& g^{ij}(x^0,\bx)+
\partial_n \zeta^i(x^0,\bx) g^{nj}(x^0,\bx)
+g^{in}(x^0,\bx)
\partial_n \zeta^j(x^0,\bx)
 \ . \nonumber \\
\end{eqnarray}
Then it is easy to see that $V_\alpha^i$
 transform under (\ref{deffpdts}) as
\begin{eqnarray}
V'^i_\tau(\tau,\sigma) &=&
V^i_\tau(\tau,\sigma)+\partial_j
\xi^i(\tau,\sigma)V^j_\tau(\tau,\sigma) \ .
\nonumber \\
V'^i_\sigma(\tau,\sigma)
&=&V^i_\sigma(\tau,\sigma)+\partial_j \xi^i(\tau,\sigma)
 V_\sigma^j(\tau,\sigma) \ .
\nonumber \\
\end{eqnarray}
Performing the same analysis as in
\cite{Kluson:2010aw}
we find that the
 Lagrangian density (\ref{mLLBS})
is invariant under target-space
foliation preserving diffeomorphism
(\ref{deffpdts}).

As the next step we check the
invariance of the action under
world-sheet foliation preserving
diffeomorphism that we define
as the world-sheet transformation
\begin{equation}\label{wsfpd}
\tau'=\tau+f(\tau) \ , \quad
\sigma'=\sigma+\epsilon(\tau,\sigma) \ .
\end{equation}
where $f,\epsilon$ are infinitesimal
parameters.
In the same way as in
\cite{Horava:2008ih} we find that
the world-sheet  metric components transform
under (\ref{wsfpd}) as
\begin{eqnarray}
n'_\sigma(\tau',\sigma')&=&
n_\sigma(\tau,\sigma)
-n_\sigma(\tau,\sigma)\parts
\epsilon(\tau,\sigma)
 -\partt f(\tau)
n_\sigma(\tau,\sigma)
 -\partt \epsilon(\tau,\sigma) \omega(\tau,\sigma)
\ ,
\nonumber \\
n'_\tau(\tau',\sigma')&=&
n_\tau(\tau,\sigma)-n_\tau(\tau,\sigma)
\partt f(\tau) \ ,
\nonumber \\
\omega'(\tau',\sigma')&=&
\omega(\tau,\sigma) -2\parts
\epsilon(\tau,\sigma)
\omega(\tau,\sigma) \ ,
\nonumber \\
n'^\sigma(\tau',\sigma')&=&
n^\sigma(\tau,\sigma)
+n^\sigma(\tau,\sigma)
\parts \epsilon(\tau,\sigma)
-n^\sigma(\tau,\sigma)
\partt f(\tau)-
\partt \epsilon(\tau,\sigma) \ .
\nonumber \\
\end{eqnarray}
Then it is easy to see that
\begin{eqnarray}
d\tau' d\sigma' n'_\tau\sqrt{\omega'}
= d\tau d\sigma n_\tau
\sqrt{\omega} \ .
\nonumber \\
\end{eqnarray}
Note that $\Gamma$ transforms under the
world-sheet foliation preserving
diffeomorphism (\ref{wsfpd})  as
\begin{eqnarray}
\Gamma'(\tau',\sigma')=
\Gamma(\tau,\sigma)-\Gamma(\tau,\sigma)
\parts \epsilon(\tau,\sigma)-
\parts^2\epsilon(\tau,\sigma) \ .
\nonumber \\
\end{eqnarray}
Then after some algebra we find that
$K_\sigma$ transforms as
\begin{eqnarray}
K_\sigma'(\tau',\sigma')=
K_\sigma(\tau,\sigma)-
2K_\sigma (\tau,\sigma)
\parts \epsilon(\tau,\sigma) \ .  \nonumber \\
\end{eqnarray}
Clearly, the world-sheet
modes $x^M$ are
scalars under (\ref{wsfpd})
\begin{equation}
x'^M(\tau',\sigma')=
x^M(\tau,\sigma) \  .
\end{equation}
Collecting all these results together
and performing the same analysis as in
\cite{Kluson:2010aw}
we can show that the
Lagrangian density
(\ref{mLLBS}) is  invariant under
world-sheet foliation preserving diffeomorphism
(\ref{wsfpd}).
\section{T-duality for LBS theory}\label{third}
In this section we analyze properties of
LBS theory under T-duality transformations.
In other words we would like to see whether
this theory shares the same properties
as ordinary string theory action.
For further purposes we again write the Lagrangian
density for LBS theory
\begin{eqnarray}\label{mlABT}
\mL&=&
-\frac{\sqrt{\omega}}{4\pi\alpha'}\frac{1}{n_\tau}
(\partt x^0-n^\sigma\parts
x^0)^2-n_\tau \sqrt{\omega} G\left(
-\frac{1}{2\pi\alpha'\omega} N^2\parts x^0
\parts x^0\right)+
\nonumber \\
&+&\omega n_\tau \frac{1}{B}
\left(\frac{1}{4\pi\alpha'}\frac{1}{n^2_\tau}
(V^i_\tau-n^\sigma V^i_\sigma)
 h_{ij}(V^j_\tau-n^\sigma V^j_\sigma)
 +\frac{1}{2(z-1)}K_\sigma \frac{1}{\omega^2}K_\sigma\right)-
 \nonumber \\
 &-& Bn_\tau\left(\frac{1}{2\pi\alpha'\omega}
 V^i_\sigma h_{ij}V^j_\sigma-A\right)-\sqrt{\omega} n_\tau F(A) \ ,
 \nonumber \\
\end{eqnarray}
where
\begin{equation}
V_\tau^i=\partt x^i+N^i\partt x^0 \ ,
\quad
V_\sigma^i=\parts x^i+N^i\parts x^0 \ .
\end{equation}
Let us now presume that
that the background
possesses isometry along $\phi$ direction
 where we performed the splitting of
the target space  coordinates
$x^i=(x^\alpha,\phi),  \alpha,\beta=1,\dots,D-1$.
The fact that there is an isometry of the background
along $\phi$ direction implies that
 the action is invariant under the
shift
\begin{equation}
\phi'(\tau,\sigma)
=\phi(\tau,\sigma)+\epsilon \ ,
\end{equation}
where $\epsilon=\mathrm{const}$.
The invariance of the action implies
an existence of the conserved
current
\begin{equation}
\mathcal{J}^\alpha
=\frac{\delta S}{\delta \partial_\alpha \phi}
 \ , \quad   \partial_\alpha \mathcal{J}^\alpha=0 \ .
 \end{equation}
explicitly, we find
\begin{eqnarray}
\mathcal{J}_\tau&=&
\frac{\omega}{2\pi\alpha'n_\tau B}h_{\phi i}
(V^i_\tau-n^\sigma V^i_\sigma) \ ,
\nonumber \\
\mathcal{J}_\sigma&=&
-\frac{ n_\sigma}{2\pi\alpha'n_\tau B}h_{\phi i}
(V^i_\tau-n^\sigma V^i_\sigma)-
B\frac{n_\tau}{\pi\alpha'\omega}h_{\phi i}V^i_\sigma
 \ .
\nonumber \\
\end{eqnarray}
Let us now try to implement the T-duality rules
as in standard bosonic theory. We gauge the
shift symmetry so that $\epsilon\rightarrow
\epsilon(\tau,\sigma)$. Then in order to
ensure the invariance of the Lagrangian
(\ref{mlABT})
we have to introduce the gauge field $a_\alpha$
and replace
\begin{equation}
\partial_\alpha\phi\rightarrow D_\alpha\phi
=\partial_\alpha\phi+a_\alpha \ .
\end{equation}
Note that under $a_\alpha$ transforms for
non-constant $\epsilon $ as
\begin{equation}
a_\alpha'(\tau,\sigma)=a_\alpha(\tau,\sigma)-
\partial_\alpha \epsilon(\tau,\sigma) \ .
\end{equation}
Then it is easy to see that
\begin{equation}
(D_\alpha \phi)'=D_\alpha \phi \ .
\end{equation}
In the same way we perform the replacement
\begin{eqnarray}
V_\tau^\phi &=&\partt \phi+N^\phi\partt x^0
\rightarrow D_\tau \phi+N^\phi\partt x^0\equiv \tV_\tau^\phi \ ,
\nonumber \\
V_\sigma^\phi&=&\parts \phi+N^\phi\parts x^0
\rightarrow
D_\sigma \phi+N^\phi\parts x^0\equiv \tV_\sigma^\phi \ .
\nonumber \\
\end{eqnarray}
However we have to also check that terms containing
$a_\alpha$ are invariant under world-sheet
foliation preserving diffeomorphism (\ref{wsfpd}).
To do this we presume that $a_\alpha$ transform
under world-sheet foliation preserving
diffeomorphism (\ref{wsfpd}) as
\begin{eqnarray}
a'_\tau(\tau',\sigma')&=&
a_\tau(\tau,\sigma)-a_\tau(\tau,\sigma)
\dot{f}(\tau)-a_\sigma(\tau,\sigma)
\partt \xi(\tau,\sigma) \ , \nonumber \\
a'_\sigma(\tau',\sigma')&=&
a_\sigma(\tau,\sigma)-a_\sigma(\tau,\sigma)
\parts \xi(\tau,\sigma) \ .
\nonumber \\
\end{eqnarray}
Then it is easy to see that the  covariant derivatives transform
as \begin{eqnarray}
D'_\tau\phi(\tau',\sigma')&=&
D_\tau\phi(\tau,\sigma)-D_\tau\phi(\tau,\sigma)
\dot{f}(\tau)-D_\sigma\phi(\tau,\sigma)
\partt \xi(\tau,\sigma) \ , \nonumber \\
D'_\sigma\phi(\tau',\sigma')&=&
D_\sigma\phi(\tau,\sigma)-D_\sigma \phi(\tau,\sigma)
\parts \xi(\tau,\sigma) \ . \nonumber \\
\end{eqnarray}
As the next step we introduce $f_{\alpha\beta}$
defined as
\begin{equation}
f_{\tau\sigma}=\partt a_\sigma-
\parts a_\tau
\end{equation}
that transform under foliation preserving diffeomorphism
as
\begin{eqnarray}
f'_{\tau\sigma}(\tau',\sigma')=
%
f_{\tau\sigma}(\tau,\sigma)
-f_{\tau\sigma}(\tau,\sigma)\dot{f}(\tau)
-f_{\tau\sigma}(\tau,\sigma)
\parts \xi(\tau,\sigma) \ .
\nonumber \\
\end{eqnarray}
Then it is easy to see that $d\tau d\sigma
 f_{\alpha\beta}$ is invariant under
(\ref{wsfpd}).
Collecting all these terms
together we  find following
Lagrangian density invariant under
the foliation preserving diffeomorphism
(\ref{wsfpd})
\begin{eqnarray}\label{tildeL}
\tilde{\mL}&=&-\frac{\sqrt{\omega}n_\tau}{4\pi\alpha'}\frac{1}{n_\tau^2}
(\partt x^0-n^\sigma\parts
x^0)^2-n_\tau \sqrt{\omega} G\left(-\frac{1}{2\pi\alpha'\omega} N^2\parts x^0
\parts x^0\right)+
\nonumber \\
&+&\omega n_\tau\frac{1}{B}
\left[\frac{1}{4\pi\alpha'n^2_\tau}
(\tV^\phi_\tau-n^\sigma \tV^\phi_\sigma)
 h_{\phi\phi}(V^\phi_\tau-n^\sigma V^\phi_\sigma)
+\frac{2}{4\pi\alpha'n^2_\tau}
(\tV^\phi_\tau-n^\sigma \tV^\phi_\sigma)
 h_{\phi \alpha}(V^\alpha_\tau-n^\sigma V^\alpha_\sigma)+\right.
 \nonumber \\
&+& \left.\frac{1}{4\pi\alpha'n^2_\tau}(V^\alpha_\tau-n^\sigma V^\alpha_\sigma)
 h_{\alpha\beta}(V^\beta_\tau-n^\sigma V^\beta_\sigma)
 +\frac{1}{2(z-1)}K_\sigma \frac{1}{\omega^2}K_\sigma\right]
  -\nonumber \\
 &-&Bn_\tau\left(\frac{1}{2\pi\alpha'\omega}
 \left[\tV^\phi_\sigma h_{\phi\phi}\tV^\phi_\sigma
 +2\tV^\phi_\sigma h_{\phi j}V^j_\sigma+
 V^\alpha_\sigma h_{\alpha\beta}V^\beta_\sigma\right]
  -A\right)-\sqrt{\omega} n_\tau F(A)
 +\tphi f_{\tau\sigma}   \ ,
 \nonumber \\
\end{eqnarray}
where $\tphi$ is the Lagrange multiplier that ensures
that $a_\alpha$ is non-dynamical field. Note
that $\tphi$ transforms as scalar under (\ref{wsfpd}).
The   gauge  invariance of the Lagrangian
density (\ref{tildeL}) can be  fixed by
imposing the condition $\phi=0$ that
implies
\begin{equation}\label{Dtau}
 D_\tau\phi=a_\tau \ , \quad
D_\sigma\phi=a_\sigma \ .
\end{equation}
Inserting (\ref{Dtau}) into (\ref{tildeL})
we obtain
\begin{eqnarray}\label{Lhat}
\mL&=&-\frac{\sqrt{\omega}}{4\pi\alpha'}\frac{1}{n_\tau}
(\partt x^0-n^\sigma\parts
x^0)^2-n_\tau \sqrt{\omega} G\left(-\frac{1}{2\pi\alpha'\omega} N^2\parts x^0
\parts x^0\right)+
\nonumber \\
&+&\frac{\omega n_\tau}{B}
\left[
\frac{1}{4\pi\alpha'n_\tau^2}
(\hV^\phi_\tau-n^\sigma \hV^\phi_\sigma)
 h_{\phi\phi}(\hV^\phi_\tau-n^\sigma \hV^\phi_\sigma)
+2(\hV^\phi_\tau-n^\sigma \hV^\phi_\sigma)
 h_{\phi \alpha}(V^\alpha_\tau-n^\sigma V^\alpha_\sigma)+\right.
 \nonumber \\
&+& \left.
\frac{1}{4\pi\alpha'n_\tau^2}
(V^\alpha_\tau-n^\sigma V^\alpha_\sigma)
 h_{\alpha\beta}(V^\beta_\tau-n^\sigma V^\beta_\sigma)
 +\frac{1}{2(z-1)}K_\sigma \frac{1}{\omega^2}K_\sigma\right]
-\nonumber \\
 &-& Bn_\tau\left(\frac{1}{2\pi\alpha'\omega}
 [\hV^\phi_\sigma h_{\phi\phi}\hV^\phi_\sigma
 +2\hV^\phi_\sigma h_{\phi \alpha}V^\alpha_\sigma+
 V^\alpha_\sigma h_{\alpha\beta}V^\beta_\sigma]
  -A\right)-\sqrt{\omega} n_\tau F(A)
 +\tphi f_{\tau\sigma}
  \ ,
 \nonumber \\
\end{eqnarray}
where
\begin{equation}
\hat{V}^\phi_\tau=
a_\tau+N^\phi\partt x^0 \ ,
\quad
\hat{V}^\phi_\sigma=
a_\sigma+N^\phi\partt x^0 \ .
\end{equation}
Now it ie easy to see that the
equation of motion  for  $\tilde{\phi}$
implies
\begin{equation}
f_{\tau\sigma}=0
\end{equation}
that can be solved as
\begin{equation}
 a_\tau=\partt \theta \ , \quad
a_\sigma=\parts \theta \ .
\end{equation}
Inserting this result into (\ref{Lhat}) we
recovery the original Lagrangian density
after replacement $\theta\rightarrow \phi$.
However the equations of motion for $a_\alpha$
that follow from (\ref{Lhat})
take more complicated  form
\begin{eqnarray}
& &\frac{\omega}{2\pi\alpha'B n_\tau}
h_{\phi\phi}(\hV^\phi_\tau-n^\sigma \hV^\phi_\sigma)
+\frac{\omega}{2\pi\alpha'B n_\tau}
h_{\phi\alpha}(\hV_\tau^\alpha-n^\sigma \hV_\sigma^\alpha)
+\parts \tphi=0 \ ,
\nonumber \\
& &-\frac{\omega n^\sigma}{2\pi\alpha'B n_\tau}
h_{\phi\phi}(\hV^\phi_\tau-n^\sigma \hV^\phi_\sigma)
-\frac{\omega n^\sigma}{2\pi\alpha' B n_\tau}
h_{\phi\alpha}(\hV^\alpha-n^\sigma \hV^\alpha_\sigma)
-\nonumber \\
&& -\frac{B n_\tau}{\pi\alpha'\omega}
(h_{\phi\phi}\hV_\sigma^\phi+h_{\phi\alpha} V^\alpha_\sigma)
-\partt \tphi=0 \ .
\nonumber \\
\end{eqnarray}
Solving these equations for
$a_\alpha$ we find
\begin{eqnarray}
a_\sigma
&=&-\frac{\pi \alpha'\omega}{h_{\phi\phi}
B n_\tau}(\partt \tphi-n^\sigma \parts \tphi)
-\frac{1}{h_{\phi\phi}}
(h_{\phi\phi}N^\phi \parts x^0+h_{\phi\alpha}
V^\alpha_\sigma) \ ,  \nonumber \\
a_\tau&=&
-\frac{n_\sigma\pi\alpha'}{Bh_{\phi\phi}n_\tau}
(\partt \tphi-n^\sigma \parts\tphi)-
\frac{2\pi\alpha'B}{\omega h_{\phi\phi}}
n_\tau \parts \tphi
-N^\phi\partt x^0-\frac{h_{\phi\alpha}}{h_{\phi\alpha}}
V^\alpha_\tau \ .  \nonumber \\
\end{eqnarray}
Inserting these results into  the Lagrangian
 density (\ref{Lhat}) we obtain the Lagrangian
 density for T-dual theory
\begin{eqnarray}\label{LagTdual}
\mL&=&-\frac{\sqrt{\omega}n_\tau}{4\pi\alpha'}\frac{1}{n_\tau^2}
(\partt x^0-n^\sigma\parts
x^0)^2-n_\tau \sqrt{\omega}
G\left(-\frac{1}{2\pi\alpha'\omega^2}N^2
\parts x^0 \parts x^0\right)+\nonumber \\
&+&\frac{\omega n_\tau }{B}
\left[\frac{1}{4\pi\alpha'n^2_\tau }
(\bV^i_\tau-n^\sigma \bV^i_\sigma)
 \tilde{h}_{ij}(\bV^j_\tau-n^\sigma \bV^j_\sigma)
+\frac{1}{2(z-1)}K_\sigma \frac{1}{\omega^2}
K_\sigma\right]-
\nonumber \\
&-& Bn_\tau\left(\frac{1}{2\pi\alpha'\omega}
 \bV^i_\sigma \hat{h}_{ij}\bV^j_\sigma-A\right)-\sqrt{\omega} n_\tau F(A)+
\nonumber \\
&+&\frac{1}{\sqrt{2}\pi\alpha'}
N^\phi(\parts x^0 \partt \tphi-
\partt x^0\parts \tphi)+
\frac{1}{\sqrt{2}
\pi\alpha'}
\frac{h_{\phi\alpha}}{h_{\phi\phi}}
(\bV^\alpha_\sigma \partt \tphi-\bV^\alpha_\tau \parts \tphi) \ ,
\nonumber \\
\end{eqnarray}
where
\begin{eqnarray}
\bV_\alpha^\phi=\partial_\alpha\tphi+
\hat{N}^\phi\partial_\alpha x^0 \ ,
\quad  \bV_\alpha^\alpha=\partial_\alpha x^\alpha+
\hat{N}^\alpha\partial_\alpha x^0 \ .
\nonumber \\
\end{eqnarray}
Note that T-dual lapse, shift and metric
components take the form
\begin{eqnarray}\label{Tdualmet}
\hat{h}_{\alpha\beta}&=&h_{\alpha\beta}
-\frac{h_{\alpha\phi}h_{\phi \beta}}{h_{\phi\phi}} \ ,
\quad   \hat{h}_{\phi\alpha}=0 \ ,
\nonumber \\
\hN&=&N \ , \quad \hat{N}^\phi=0 \ , \quad  \hN^\alpha=N^\alpha \ . \nonumber \\
\end{eqnarray}
Note that T-dual metric written in
$D+1$ formalism takes the form
\begin{equation}
\hg_{00}=-\hN^2+\hN_i\hh^{ij}\hN_j \ , \quad
\hg_{0i}=\hN_i \ , \quad \hg_{ij}=\hh_{ij} \ ,
\end{equation}
with inverse
\begin{equation}
\hg^{00}=-\frac{1}{\hN^2} \ , \quad
\hg^{0i}=\frac{\hN^i}{\hN} \ , \quad
\hg^{ij}=\hh^{ij}-\frac{\hN^i \hN^j}{\hN^2} \ .
\end{equation}
From (\ref{Tdualmet})
we see that in T-dual theory $\hg_{\phi 0}=
\hg_{\phi \alpha}=0$
and also
\begin{equation}
\hh^{\phi\phi}=\frac{1}{\hh_{\phi\phi}} \ ,
\quad
\hh^{\phi\alpha}=0 \ .
\end{equation}
Then with the help of (\ref{Tdualmet})
we find
\begin{eqnarray}\label{gTdual}
\hg_{00}
&=&-\hN^2+\hN^\alpha \hh_{\alpha\beta}\hh^\beta+
\hN^\phi \hh_{\phi\phi}\hN^\phi
=g_{00}-\frac{g_{0\phi}g_{\phi 0}}
{g_{\phi\phi}} \ .
\nonumber \\
\end{eqnarray}
Finally we consider the last term in
(\ref{LagTdual})
\begin{eqnarray}
& &\frac{1}{\sqrt{2}\pi\alpha'}
N^\phi(\parts x^0 \partt \tphi-
\partt x^0\parts \tphi)+
\frac{h_{\phi\alpha}}{\sqrt{2}
\pi\alpha'h_{\phi\phi}}
(V^\alpha_\sigma \partt \tphi-V^\alpha_\tau \parts \tphi)=
\nonumber \\
&=&-\frac{1}{\sqrt{2}\pi\alpha'}
\frac{N_\phi}{h_{\phi\phi}}
(\partt x^0\parts \tphi-\parts x^0 \partt \tphi)-
\frac{h_{\phi\alpha}}{\sqrt{2}
\pi\alpha'h_{\phi\phi}}
(\partt x^\alpha \parts \tphi-\parts x^\alpha_\sigma \partt \tphi)=
\nonumber \\
&=&\frac{1}{\sqrt{2}\pi\alpha'}
\hat{b}_{0\phi}
(\partt x^0\parts \tphi-\parts x^0 \partt \tphi)+
\frac{1}{\sqrt{2}
\pi\alpha'}
\hat{b}_{\alpha\phi}
(\partt x^\alpha \parts \tphi-\parts x^\alpha_\sigma \partt \tphi) \ ,
\nonumber \\
\end{eqnarray}
where
\begin{equation}\label{TdualB}
\hat{b}_{0\phi}=-\frac{N_\phi}{h_{\phi\phi}}=
-\frac{g_{0\phi}}{g_{\phi\phi}} \ ,
\quad
\hat{b}_{\alpha\phi}=
-\frac{h_{\phi\alpha}}{h_{\phi\phi}}
=-\frac{g_{\phi\alpha}}{g_{\phi\phi}} \ .
\end{equation}
We see that the relations between original
and T-dual background fields given in
(\ref{Tdualmet}),(\ref{gTdual}) and (\ref{TdualB})
 exactly coincide
with the standard Buscher's rules
\cite{Buscher:1985kb,Buscher:1987sk,Buscher:1987qj} between original
and T-dual metric components (see also
\cite{Bergshoeff:1995as,Breckenridge:1996tt})
\begin{eqnarray}
\hat{g}_{\phi\phi}&=&\frac{1}{g_{\phi\phi}} \ , \quad
\hat{g}_{\phi 0}=\frac{b_{\phi 0}}{g_{\phi\phi}} \ , \quad
\hat{g}_{\phi\alpha}=\frac{b_{\phi\alpha}}{g_{\phi\phi}} \ ,
\nonumber \\
\hat{g}_{00}&=&g_{00}-\frac{g_{0\phi}g_{\phi 0}}{g_{00}} \ ,
\quad
\hat{h}_{\alpha\beta}=g_{\alpha\beta}-\frac{g_{\alpha\phi}
g_{\phi\beta}}{g_{\phi\phi}} \ , \nonumber \\
\hat{b}_{\phi 0}&=&\frac{g_{\phi 0}}{g_{\phi\phi}} \ ,
\quad \hat{b}_{0\phi}=-\frac{g_{\phi 0}}{g_{\phi\phi}} \  ,
\quad
\hat{b}_{\phi\alpha}=\frac{g_{\phi\alpha}}{g_{\phi\phi}} \ ,
\quad \hat{b}_{\alpha\phi}=-\frac{g_{\alpha\phi}}{g_{\phi\phi}} \ .
\nonumber \\
\end{eqnarray}

\vskip 5mm

 \noindent {\bf
Acknowledgements:}
J. K. would like to thank CERN PH-TH for
generous hospitality and financial
 support during the course of this work.  J.K. is
 also supported by
 the Czech Ministry of
Education  under Contract No. MSM
0021622409. \vskip 5mm

\end{document}